\documentclass[
 reprint,
 amsmath,amssymb,
 aps,
floatfix,
]{revtex4-2}

\def\aap{A\&A}
\def\apj{ApJ}

\def\apjl{ApJ}
\def\mnras{MNRAS}
\def\araa{ARA\&A}

\def\na{New Astronomy}

\def\physrep{Phys. Rep.}
\def\nat{Nature}

\def\apjs{ApJS}
\def\prd{Phys. Rev. D}

\usepackage{xspace}
\usepackage{subcaption}
\usepackage{multirow}
\usepackage{graphicx}
\usepackage{dcolumn}
\usepackage{bm}
\usepackage{hyperref}
\hypersetup{
	colorlinks=true,        
	linkcolor=blue,        
	citecolor=blue,        
	urlcolor=blue,           
	}
\usepackage{dsfont}

\usepackage{comment}
\usepackage{color}
\usepackage[normalem]{ulem}
\usepackage{soul}
\usepackage{float}

\begin{document}

\preprint{APS/123-QED}

\title{Effect of quark-hadron phase transitions on the neutron star mass distribution}

\author{Renyu Luo$^{1}$}
\author{Xu-Run Huang$^{2}$}
\author{Shuai Zha$^{3}$}
\author{Guoliang L\"{u}$^{1}$}
\email{guolianglv@sina.com}
\author{Chunhua Zhu$^{1}$}
\author{Sufen Guo$^{1}$}
\author{Helei Liu$^{1}$}
\author{Zhijun Wang$^{1}$}

\affiliation{
 $^{1}$School of Physical Science and Technology, Xinjiang University, Urumqi, 830046, China\\
 $^{2}$Department of Physics, Chongqing Key Laboratory for Strongly Coupled Physics, Chongqing University, Chongqing 401331, China\\
 $^{3}$Yunnan Observatories, Chinese Academy of Sciences, Kunming, 650216, China\\
}

\date{\today}

\begin{abstract}
The neutron star (NS) mass distribution serves as a crucial probe of the dense matter equation of state (EOS) and core-collapse supernova explosion mechanisms. Although observations reveal a bimodal NS mass distribution, existing theoretical simulations often struggle to naturally reproduce the high-mass peak, and the impact of the quark-hadron phase transition (PT) in high-density nuclear matter remains largely unexplored. Employing the 1D general-relativistic hydrodynamics code GR1D with hybrid EOSs incorporating a first-order PT (RDF1.1 and RDF1.2), we systematically simulate core-collapse supernovae for progenitors with zero-age main-sequence masses ranging from 13 to 120 $M_\odot$. Our results reveal a threefold impact of the PT onset density on the final supernova fate: in low-compactness progenitors, an early PT truncates accretion and produces less massive NSs; in the critical-compactness regime, it facilitates successful explosions; and in high-compactness progenitors, it accelerates black hole formation. Compared with the $\sim 1.4\,M_\odot$ peak typical of the conventional core-collapse scenario, the inclusion of the PT produces NSs with masses centered around $1.70\,M_\odot$ for RDF1.1 and within $1.55$--$1.70\,M_\odot$ for RDF1.2. Furthermore, we identify a novel formation channel whereby low-compactness progenitors with failed initial shocks can form more massive NSs via a PT induced second shock. Due to the fixed nature of the PT boundary, the final NS masses exhibit a pronounced clustering effect. This study elucidates the critical role of the quark-hadron PT in shaping the NS mass distribution, providing important astrophysical constraints on the properties of high-density nuclear matter and the formation mechanism of massive NSs.
\end{abstract}

\maketitle

\section{Introduction}

Neutron stars (NSs), the end products of massive stellar evolution, have mass distributions that provide critical probes of supernova collapse, compact binary evolution, and the dense matter equation of state (EOS) \citep{2015Prama..84..927P,2016PhR...621..127L,2025RvMP...97d5007C,2025PhRvD.112j3005L}. Since the first pulsar discovery in 1967 \citep{1968SciAm.219d..25H}, NS mass measurements have steadily improved in precision and sample size, advancing our understanding of their formation and evolution. \citet{1994PhRvL..73.1878F} performed the first Bayesian analysis of NS mass distributions, analyzing eight measurements from four double NS systems and concluding that masses cluster predominantly in 1.3--1.6 $M_{\odot}$. \citet{1999ApJ...512..288T} subsequently expanded the sample to 19 objects, finding a mean mass of $1.35 \pm 0.04\,M_{\odot}$ with negligible dispersion and solidifying the single formation channel paradigm. Although \citet{2001A&A...377..925B} discovered Vela X-1 ($1.86 \pm 0.16\,M_{\odot}$), the first massive NS found to deviate significantly from the canonical 1.4 $M_{\odot}$ this outlier did not challenge the classical paradigm due to the limited sample size and measurement precision of the time. Many more such massive NSs have since been observed \citep{2005ApJ...634.1242N,2008ApJ...679.1433F,2008ApJ...675..670F,2010Natur.467.1081D}. In 2010, \citet{2010ApJ...719..722S} first reported a possible bimodal NS mass distribution using a high-precision sample (uncertainties $< 0.025\,M_{\odot}$): a main peak at 1.35 $M_{\odot}$ from standard iron core-collapse supernovae (CCSNe), and a secondary peak at $\sim 1.25\,M_{\odot}$ attributed to electron-capture supernovae of oxygen-neon-magnesium cores \citep{1984ApJ...277..791N,2004ApJ...612.1044P,2019ApJ...886...22Z,2022MNRAS.513.1317Z}. Subsequent large-scale statistical studies confirmed the ubiquity of this bimodality but placed the second peak in the higher $1.75-1.8\,M_{\odot}$ range \citep{2011MNRAS.414.1427V,2013ApJ...778...66K,2016ARA&A..54..401O,2024Univ...10....3R}.

Theoretical predictions of NS mass distributions rely critically on the understanding of the explosion mechanism of CCSNe from massive stars \citep{2012MNRAS.424.1570P,2021Natur.589...29B}. In 1996, \citet{1996ApJ...457..834T} first predicted, based on the piston explosion model \citep{1995ApJS..101..181W}, that the NS birth mass function should exhibit a bimodal distribution with peaks at 1.27 $M_\odot$ and 1.76 $M_\odot$. The underlying physical mechanism lies in the pre-supernova structure bifurcation: stars with main-sequence masses below $\sim$19 $M_\odot$ undergo convective core carbon burning, producing low central entropy, smaller iron cores, and ultimately low-mass NSs; those above this threshold burn carbon radiatively in their cores, leading to significantly higher central entropy, forming larger iron cores, and yielding more massive NSs \citep{1996ApJ...457..834T,1993PhR...227...65W}. This theoretical prediction preceded the observational confirmation of bimodality. However, subsequent studies based on more sophisticated one-dimensional or semi-analytical models \citep{2012ApJ...757...69U,2012ApJ...749...91F} by \citet{2016ApJ...821...38S}, \citet{2018ApJ...856...35R}, \citet{2020MNRAS.499.3214M}, \citet{2022ApJ...932L...3M} and \citet{2022ApJ...931...94F} generally predicted that the NS birth mass distribution should be unimodal, with a peak around $1.2-1.4\,M_{\odot}$. These studies generally argued that high-compactness progenitors (the high-entropy stars that \citet{1996ApJ...457..834T} had predicted would form 1.73--1.8 $M_{\odot}$ NSs) could not produce successful explosions and would instead collapse directly to black holes, thus preventing the formation of a second peak in the mass distribution. However, multidimensional studies have demonstrated that multidimensional effects, such as neutrino driven turbulence and convection  \cite{2008ApJ...688.1159M,2018ApJ...856...22M}, can facilitate the explosions of high-compactness pre-supernova progenitors \cite{2020MNRAS.491.2715B,2022MNRAS.517..543W,2024ApJ...964L..16B,2023ApJ...952...79L}. Through 3D simulations of twenty pre-supernova models, \citet{2024ApJ...964L..16B} found a robust, approximately monotonic correlation between the compactness and the neutron star mass. By leveraging this relationship, along with the correlation between the zero-age main-sequence (ZAMS) mass and the compactness, we can derive the birth mass function for NSs. \citet{2024PhRvD.110b3007B} adopted a 1D+ turbulence model (a model that parametrizes 2D/3D neutrino-driven turbulent convection effects via mixing-length theory \citep{2020ApJ...890..127C,2021ApJ...912...29B}, with the mixing-length parameter $\alpha_{\text{MLT}}$ calibrated against three-dimensional simulations \citep{2021ApJ...912...29B,2022ApJ...926..147B} and observational constraints on the fraction of stars that collapse to black holes \citep{2023ApJ...949...17B}) to simulate 341 pre-supernova progenitor models, and successfully reproduced a bimodal NS mass distribution. This study argues that these high-compactness progenitors can indeed undergo successful explosions through multidimensional effects such as neutrino-driven convection and turbulence, thereby forming the second peak. Nevertheless, the recent work by \citet{2026arXiv260516504G} indicates that, when neutrino flavor conversion (FC) is taken into account, the high-compactness progenitors that form the second peak in the NS mass distribution in \citet{2024PhRvD.110b3007B} may fail to explode successfully. This ongoing theoretical debate highlights that the existence of the second peak in the single-star NS birth mass distribution remains highly sensitive to the microphysical processes incorporated in CCSNe simulations \citep{2025PhRvD.111j3004L}.

To date, all NSs with precisely measured masses derived from dynamical methods reside in binary systems \citep{1999ApJ...512..288T,2016ARA&A..54..401O}. In close binaries, NSs can gain mass by accreting material from their companions via Roche-lobe overflow, for which the existence of millisecond pulsars serves as direct evidence of this recycling process \citep{1982Natur.300..728A,1982CSci...51.1096R,1991PhR...203....1B,2015MNRAS.454.1725Z}. However, the accretion process onto NSs is governed by several complex physical mechanisms \citep{1996ApJ...464L.139V,2011MNRAS.416.2130T,2011MNRAS.416.2285L,2018NewA...62...94R}. Early models, constrained solely by the Eddington limit, may overestimate the accreted mass \citep{1999A&A...350..928T,2002ApJ...565.1107P,2011ApJ...732...70L,2019ApJ...886L..31V}; whereas \citet{2021ApJ...922..158L}, by further accounting for matter ejection due to the propeller effect, estimated a relatively limited accreted mass increment for NSs during the recycling process. In recent years, observational and theoretical studies have shown that NSs can achieve significant mass growth through accretion in spider systems \citep{2020SCPMA..6329531H,2023ApJ...942....6K,2024Univ...10....3R} or via super-Eddington accretion \citep{2014Natur.514..202B,2012MNRAS.424.2265L,2026arXiv260524917H}. Recently, \citet{2025NatAs...9..552Y} performed a systematic statistical analysis to recover the birth mass function by applying probabilistic corrections for accreted mass to a large neutron star sample. Accounting for the diverse accretion processes discussed above, they inferred a unimodal distribution for birth masses, suggesting that the empirical bimodal distribution observed in NS masses is likely a consequence of binary accretion history. Consequently, binary accretion serves as another important avenue for understanding the origin of the NS mass distribution. 

Notably, none of the prior theoretical investigations into the NS mass distribution have systematically incorporated the potential impacts of the hadron-quark phase transition (PT) in high-density nuclear matter. During the postbounce accretion phase, the central density of the proto-neutron star (PNS) can reaches several times the nuclear saturation density ($\rho_{\mathrm{sat}}$) \citep{2017hsn..book.1095J,2023arXiv230409350H}. At such densities, free quarks may become deconfined from hadrons \citep{1984PhRvD..30..272W}, and the central region of the PNS enters the hadron-quark mixed phase \citep{2009PhRvL.102h1101S,2021ApJ...911...74Z,2025ApJ...979..151H}. As the quark matter fraction increases, the central adiabatic index of the PNS decreases, leading to a second collapse \citep{2009PhRvL.102h1101S}. The collapse continues until the formation of the pure quark phase, which stiffens the EOS and halts the collapse \citep{2009PhRvL.102h1101S}. Similar to the first shock, the rapidly infalling matter rebounds upon impacting the central high-density core, launching a second shock wave that helps to trigger the supernova explosion, and simultaneously releasing a large burst of electron antineutrinos \citep{2009PhRvL.102h1101S,2020PhRvL.125e1102Z,2021ApJ...911...74Z}. This PT explosion mechanism not only provides a plausible explanation for the explosions of massive stars ( $\sim$ 50 $M_{\odot}$ blue supergiants), but crucially, it directly produces hybrid stars (NSs with a quark matter core) with a birth mass of $\sim$ 2 $M_{\odot}$, thereby offering a potential origin for the high-mass tail of the observed NS mass distribution \citep{2018NatAs...2..980F}. However, how the PT systematically affects the structure of the bimodal NS mass distribution remains unclear. To this end, we employed an EOS with PT and used the 1D CCSNe simulation code GR1D \citep{2010CQGra..27k4103O,2015ApJS..219...24O} to perform systematic numerical simulations of a suite of pre-supernova models with different initial masses, statistically derived the NS mass distribution, and thereby clarified the effect of PT on the NS mass distribution, filling the gap in systematic studies in this research direction.

The primary focus of this work is to investigate the effect of PT on the NS mass distribution. In Section \ref{sec:methods}, we describe our model setup and the methods employed. Our results are presented in Section \ref{sec:results}, followed by a summary in Section \ref{subsec:Conclusion}.

\section{Methods}\label{sec:methods}

In this section, we present the model parameter settings for the GR1D code. We also describe the criteria for NS mass determination adopted in this work.
\begin{figure*}[htbp]
    \includegraphics[width=1.\textwidth]{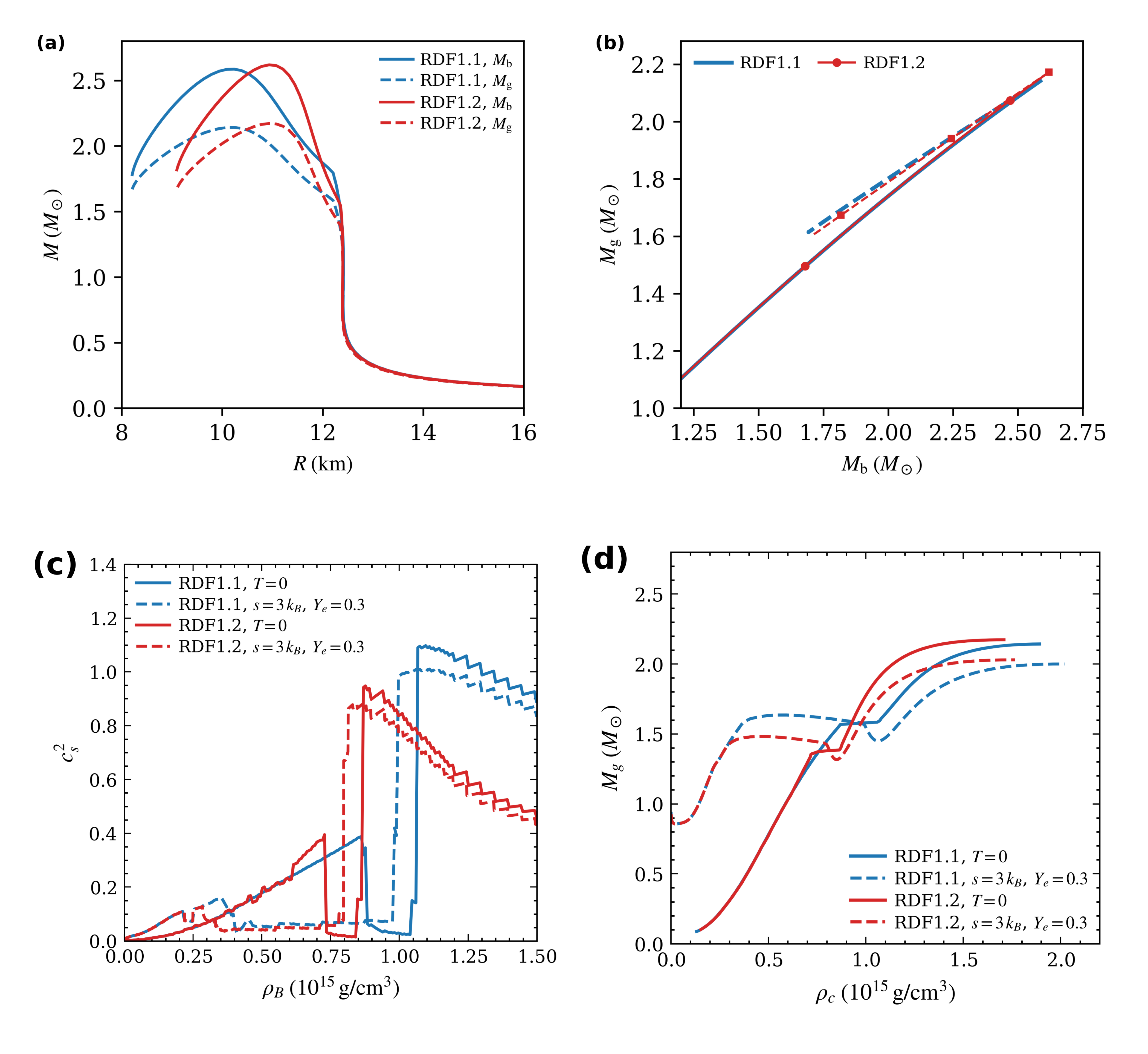}
    \caption{Properties of compact stars described by the RDF1.1 (blue) and RDF1.2 (red) EOSs. (a) Cold mass-radius relations from the TOV equations (solid:$M_b$--$R$; dashed: $M_g$--$R$). (b) Cold baryonic versus gravitational mass relations; the solid lines (stable branches) are used for mass conversion. (c) Squared sound speed $c_s^2$ versus baryon density $\rho_B$ and (d) gravitational mass $M_g$ versus central baryon density $\rho_c$, at zero temperature (solid) and at $s = 3\,k_B$, $Y_e = 0.3$ (dashed). }
    \label{fig:1}
\end{figure*}
\subsection{Core-collapse model}

There are many pre-supernova progenitor models (\citet{2002RvMP...74.1015W,2003ApJ...591..288H,2016ApJ...821...38S,2018ApJ...860...93S,2023ApJ...952...79L}). We adopt the solar-metallicity stellar evolution models of \citet{2016ApJ...821...38S}, which provide a dense grid in ZAMS mass together with the corresponding CCSNe outcomes (namely, explodability and compact remnant mass) for each progenitor, facilitating a direct comparison with the present work. We select models with ZAMS masses in the range $13$ to $120\,M_\odot$, sampling the $13$--$30\,M_\odot$ interval at a uniform spacing of $0.2\,M_\odot$. The core collapse and explosion phases are simulated using the GR1D code, which employs the M1 scheme for neutrino transport \citep{2015ApJS..219...24O}. The M1 transport is a two-moment scheme that solves the Boltzmann equation governing neutrino propagation \citep{2011PThPh.125.1255S,2013PhRvD..87j3004C}. The computational grid consists of 600 radial zones, with a uniform spacing of 0.3 km for the inner 20 km. Beyond 20 km, the grid extends logarithmically to an outer boundary where the density drops to 8000 g cm$^{-3}$. 

For the hybrid EOS incorporating a first-order PT, we adopt the RDF1.1 and RDF1.2 models (Fig~\ref{fig:1}(a)) \citep{2021PhRvD.103b3001B}. The hadronic phase is based on the DD2 \citep{2010PhRvC..81a5803T} relativistic mean-field model: RDF1.1 employs the DD2F model corrected for the nuclear matter flow constraint \citep{2002Sci...298.1592D,2016EPJA...52...69A}, whereas RDF1.2 utilizes DD2Fev, which further incorporates an excluded-volume modification \citep{2010NuPhA.837..210H}. The quark phase is described by the relativistic density functional (RDF) quark model \citep{2021PhRvD.103b3001B,2017PhRvD..96e6024K}, featuring a string-flip mechanism to mimic quark confinement \citep{1985PhRvD..31.1689H,1986PhRvD..34.3499R} and linear plus higher-order vector repulsions \citep{2021PhRvD.103b3001B} to satisfy the observational constraint of $\sim$ 2 $M_{\odot}$ NSs \citep{2010Natur.467.1081D,2013Sci...340..448A}.  Both EOSs are constructed via the Gibbs construction for the PT \citep{2021PhRvD.103b3001B}. Compared to RDF1.1, RDF1.2 has a lower onset density ($\sim 3.15 \times 10^{14} \; \mathrm{g/cm^{3}} \text{ vs. } \sim 4.07 \times 10^{14} \; \mathrm{g/cm^{3}}$) and a correspondingly lower onset mass $M_{onset}$ ($1.44\ M_{\odot}$ vs.\ $1.60\ M_{\odot}$) at entropy $s = 3\ k_{B}$ and electron fraction $Y_{e} = 0.3$ (see Fig~\ref{fig:1}(c) and \ref{fig:1}(d)), along with a narrower mixed-phase density interval \citep{2024ApJ...964..143K}, which favors the triggering of a PT-driven explosion mechanism in supernova simulations \citep{2024ApJ...964..143K,2018NatAs...2..980F}. We generated the neutrino opacity table using Nulib \citep{2015ApJS..219...24O}, the open-source code distributed with GR1D, primarily adopting default parameter settings. 

In this study, we perform two sets of simulations, adopting the RDF1.1 and RDF1.2 EOSs, respectively, without activating the Supernova Turbulence in Reduced dimensionality (STIR) model \citep{2021ApJ...912...29B,2020ApJ...890..127C}, as the necessary calibration of the mixing-length parameter $\alpha_{\text{MLT}}$ cannot be carried out. While a value of $\alpha_{\text{MLT}}$ $=$ 1.51 was calibrated by \citet{2023ApJ...949...17B} using 3D simulations with the SFHo EOS, the lack of 3D simulations for the RDF1.1 and RDF1.2 EOSs prevents us from performing a similar calibration. Given that multidimensional simulations have established the decisive role of neutrino-driven turbulent convection in CCSNe explosions \citep{2015ApJ...799....5C,2016ApJ...820...76R,2013ApJ...771...52M,2008ApJ...688.1159M,2018JPhG...45e3003R}, the inability to capture this effect with the STIR model represents a significant limitation of the present work.

\subsection{Neutron star mass determination}\label{sec:2.2}
In our simulations, the calculations are continued until the code terminates due to numerical failure. Some of the models where the PT-induced secondary shock has been successfully launched, the electron fracti $Y_{e}$ tends to exceed the upper limit permitted by EOS once the shock has propagated to approximately 500–2000 km, leading to termination of the computation. For these models, since the PT-induced secondary shock has already been successfully launched at the time of termination, we assume they would ultimately explode and include them in our sample of successful PT-induced explosions. We note, however, that this early termination may cause our simulations to miss some cases of delayed black hole formation or late time failures that would occur at later stages if the EOS permitted further evolution. This primarily affects the relative number of NSs formed via the PT channel, not the position of the peak in the resulting mass distribution, which is set by the EOS dependent PT onset conditions. Following common practice in the literature \citep{2006A&A...457..963S,2012ApJ...756...84M,2020MNRAS.496.2039S,2019MNRAS.485.3153B,2019MNRAS.485.3153B} , the baryonic mass of the PNS is defined as the baryonic mass enclosed within the region where the density exceeds $10^{11}\ \mathrm{g\ cm^{-3}} $ at the end of the simulation. According to Fig.4 of \citet{2024ApJ...964..143K}, when this mass criterion is applied, the NS mass obtained at the evolutionary stage reached in our simulations is essentially consistent with that obtained by continuing the simulation until the shock traverses the entire computational domain.

In Appendix~\ref{app:A}, we present the NS masses obtained from simulations of the s40 \citep{2002RvMP...74.1015W} progenitor model used by \citet{2024ApJ...964..143K} with various equations of state. Although \citet{2024ApJ...964..143K} employed the AGILE-BOLTZTRAN code \citep{1993ApJ...405..637M,1993ApJ...410..740M,1993ApJ...405..669M} whereas the present study uses the GR1D code, according to the benchmark comparison of these two supernova simulation codes by \citet{2018JPhG...45j4001O}, the typical variations between independent codes are within about 10\%, with many critical quantities (such as the peak neutrino heating rate and electron type neutrino luminosities) agreeing to within 5\%. As shown in Appendix~\ref{app:A}, the PNS baryonic masses obtained in this work agree with those of \citet{2024ApJ...964..143K} to within 0.02 $M_{\odot}$. This deviation lies well within the systematic uncertainties expected between different simulation codes, thereby robustly validating the reliability of our simulation results.

Our simulations directly yield the baryonic mass of the PNS (the hot NS), whereas astronomical observations provide the gravitational mass of cold NSs. Given the conservation of baryon number during the cooling phase of a NS, the baryonic mass of the hot PNS is identical to that of the cold NS \citep{1999ApJ...513..780P,1986ApJ...307..178B}. Accordingly, we calculate the conversion relation between the gravitational and baryonic masses using the adopted EOS under near-zero temperature conditions, as shown in the Fig.~\ref{fig:1}(b). For the mass conversion, we adopt the solid branches in the Fig.~\ref{fig:1}(b), which represent stable hydrostatic equilibrium configurations \cite{1983bhwd.book.....S}.

It is important to note that the NS masses derived in this study do not include the effects of post-explosion fallback accretion. Fallback accretion is intrinsically a late-time hydrodynamic phenomenon, the accurate evaluation of which requires tracking the shock propagation through the entire stellar envelope and the subsequent dynamical evolution of the ejecta at large radii. As mentioned earlier, our simulations terminate due to numerical divergence when the shock wave reaches a radius of about 500--2000 km ; at this stage, the shock is still deep inside the progenitor star. Consequently, limited by the temporal and spatial scales of our simulations, capturing the fallback accretion \citep{1989ApJ...346..847C,2008ApJ...679..639Z,2023ApJ...957...68B} process is fundamentally unfeasible. This treatment is consistent with the recent work of \citet{2024PhRvD.110b3007B}, who similarly evaluated remnant masses from 1D CCSNe simulations without accounting for late-time fallback accretion. Therefore, the masses reported in this work represent the PNS masses at the termination of our simulations, excluding any subsequent accretion increments.
\section{Results}\label{sec:results}
In this section, we present in detail the results of core-collapse supernova simulations incorporating the PT. We first simulate multiple pre-supernova models using two equations of state, RDF1.1 and RDF1.2, to analyze the influence of the phase transition on the supernova explosion process and compare the evolutionary differences between the two EOSs (Sec.~\ref{sec:3.1}). Subsequently, we present the explosion outcomes of the two sets of simulations (Sec.~\ref{sec:3.2}). Finally, we correlate the initial compactness of the pre-supernova progenitor with the final NS mass and the explosion timescale, elucidating a novel physical channel through which low compactness stars form relatively more massive NSs via a PT induced second shock (Sec.~\ref{sec:3.3}).
\subsection{Impact of different PT EOS on core-collapse outcomes}\label{sec:3.1}
\begin{figure*}[htbp]
    \includegraphics[width=1.\textwidth]{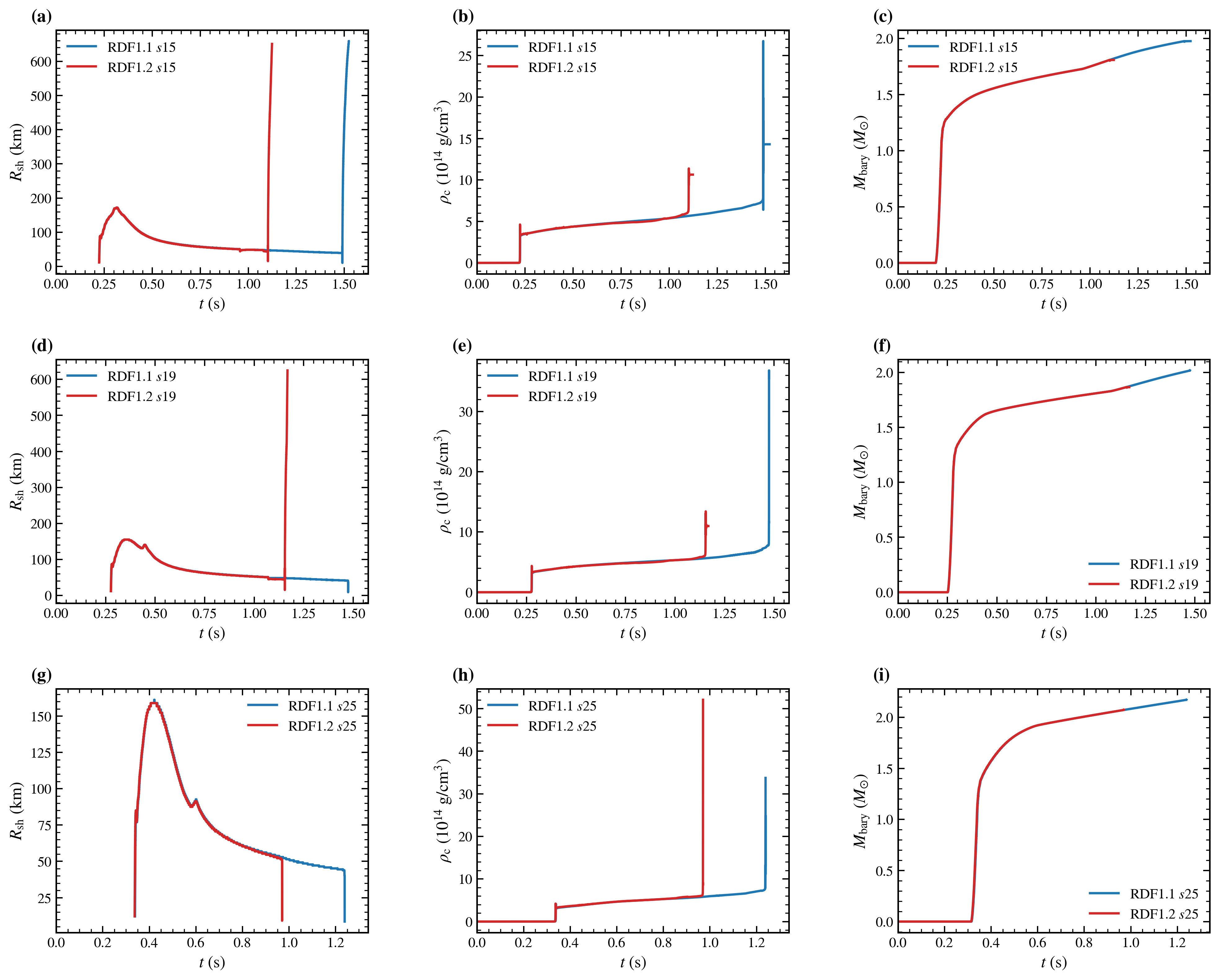}
    \caption{Time evolution of shock radius (panels a, d, g), central density (panels b, e, h), and PNS mass (defined as the baryonic mass enclosed within the $\rho = 10^{11}\ \mathrm{g\,cm^{-3}}$ contour; panels c, f, i) for CCSNe simulations with initial stellar masses of 15 $M_{\odot}$ (top panels), 19 $M_{\odot}$ (middle panels), and 25 $M_{\odot}$ (bottom panels). Blue lines denote results from simulations using the RDF1.1 equation of state, while red lines correspond to RDF1.2. }
    \label{fig:2}
\end{figure*}

To elucidate the specific effects of the PT on the dynamics of supernova explosions, we select three representative pre-supernova models with ZAMS masses of 15 $\rm{M_{\odot}}$ ($\xi_{1.75}$ = 0.60), 19 $\rm{M_{\odot}}$ ($\xi_{1.75}$ = 0.72), 25 $\rm{M_{\odot}}$ ($\xi_{1.75}$ = 1.17), and compare their evolution under two EOSs, RDF1.1 and RDF1.2. The compactness parameter is defined as \cite{2011ApJ...730...70O}:
\begin{equation}
\xi_{M} = \frac{M/M_{\odot}}{R(M_{\rm bary}=M)/(1000~\rm km)},
\end{equation}
where we set \( M = 1.75 M_{\odot} \) following \citet{2024ApJ...964L..16B}, and \( R(M) \) refers to the radius of a closed region with a baryonic mass of \( M \), which we determine directly from the pre-supernova profiles of \citet{2016ApJ...821...38S}. The three models span a wide range of compactness, covering the regimes of successful explosion, critical transition, and direct black hole formation, thereby allowing us to systematically probe the dependence of explosion outcomes on the PT onset density.

Since RDF1.1 and RDF1.2 differ only in an excluded-volume correction to the hadronic model (DD2F vs. DD2Fvex), which becomes significant only at high densities \cite{2021PhRvD.103b3001B}, the evolutionary trajectories for the two EOSs coincide exactly during the early stage of core collapse. As shown in Fig.~\ref{fig:2}(b, e, h), both EOS models exhibit the same peak central density at core bounce; meanwhile, Fig.~\ref{fig:2}(c, f, i) demonstrate that they possess nearly identical initial PNS baryonic masses. Subsequently, the initial shock rapidly loses momentum during its outward propagation owing to photodisintegration and neutrino energy losses \citep{2017hsn..book.1095J,1966ApJ...143..626C}, stalls at approximately 160 km , and enters the accretion phase [Fig.~\ref{fig:2}(a, d, g)]. During this phase, the PNS steadily gains mass through continuous accretion of the overlying material [Fig.~\ref{fig:2}(c, f, i)], while the central density rises gradually [Fig.~\ref{fig:2}(b, e, h)].

For the $15\,M_\odot$ model, once the central density of the PNS reaches the critical threshold for the PT, hadronic matter begins to convert into quark matter. Since the onset density of the PT is lower for RDF1.2 than for RDF1.1, the transition is triggered earlier, and the PNS mass at the onset of the second collapse is correspondingly smaller. As shown in Fig.~\ref{fig:2}(b), the second collapse is triggered earlier in the RDF1.2 model, resulting in a rapid increase in the central density; subsequently, the formation of a pure quark phase dramatically stiffens the EOS, inducing a strong bounce that successfully launches the second shock wave. This difference in the timing of the PT directly determines the final mass of the neutron star. Because RDF1.2 triggers the second shock earlier and successfully revives the shock [Fig.~\ref{fig:2}(a)], the outward-propagating shock effectively truncates the accretion flow, resulting in a significantly shorter effective accretion timescale for the PNS compared to RDF1.1. Consequently, although RDF1.2 undergoes the PT earlier, it ultimately produces a remnant with a baryonic mass of only $1.81\,M_\odot$ (cold gravitational NS mass $1.60\,M_\odot$), which is significantly lower than the $1.98\,M_\odot$ (cold gravitational NS mass $1.72\,M_\odot$) produced by RDF1.1 [Fig.~\ref{fig:2}(c)]. This result clearly demonstrates that in models with successful explosions, the critical density for the PT and the PNS mass at the onset of the second collapse play a decisive "truncation" role in determining the final remnant mass by dictating when the shock cuts off the accretion flow.

As the pre-supernova compactness increases, the evolution enters a highly sensitive critical regime. For the 19 $M_{\odot}$ progenitor, the two EOSs lead to distinctly different final fates. For RDF1.2, the lower onset density of the phase transition triggers the second collapse at a relatively small PNS mass. Since the PNS mass has not yet exceeded the maximum mass limit of the quark phase, the pure quark phase formed after the second collapse provides sufficient pressure support. This results in a sharp stiffening of the EOS and an effective bounce, successfully launching the second shock wave to revive the explosion [Fig.~\ref{fig:2}(d)] and leaving behind a neutron star ($M_{bary}$ about 1.87 $M_{\odot}$ and cold $M_{garv}$ about 1.64 $M_{\odot}$). In contrast, RDF1.1 requires a longer accretion time and a larger PNS mass to trigger the second collapse due to its higher onset density [Fig.~\ref{fig:2}(f)]. However, at this stage, the excessively high PNS mass exceeds the maximum mass limit supported by the quark phase. Consequently, the core fails to generate an effective bounce and collapses directly into a black hole [Fig.~\ref{fig:2}(d, e)]. This comparison reveals a crucial physical mechanism: in the critical compactness regime, an earlier phase transition (RDF1.2) ensures that the quark phase can effectively support the core and trigger an explosion, whereas a delayed transition (RDF1.1) misses the explosion window due to excessive mass accretion, resulting in direct black hole formation.

When the compactness further increases to 25 $M_{\odot}$, the extremely high mass accretion rate prevents any successful explosion, leading to black hole formation for both EOSs. For RDF1.2, the low onset density triggers the second collapse at a lower mass; however, similar to the 19 $M_{\odot}$ case with RDF1.1, the extreme accretion rate overwhelms the pressure support of the quark phase, causing immediate collapse into a black hole. RDF1.1 experiences a longer accretion phase before the phase transition but similarly collapses directly into a black hole. This process is clearly illustrated in the PNS mass evolution curves [Fig.~\ref{fig:2}(i)]: the mass curve for RDF1.2 terminates at an earlier time, while RDF1.1 undergoes a prolonged accretion phase before termination. This indicates that in models with extremely high accretion rates, a lower phase transition onset density accelerates black hole formation.

In summary, the comparison across the three progenitor models comprehensively reveals the threefold impact of the phase transition properties, most notably the onset density, on the final fate of CCSNe: in the low-compactness regime, an earlier transition shortens the accretion time, resulting in a less massive neutron star; in the critical regime, it prevents the core mass from exceeding the quark phase support limit, determining the success or failure of the explosion; and in the high-compactness regime, it accelerates black hole formation as the quark phase fails to support the core.

\subsection{Explosion results}\label{sec:3.2}
\begin{figure}[ht]
	\includegraphics[width=.5\textwidth]{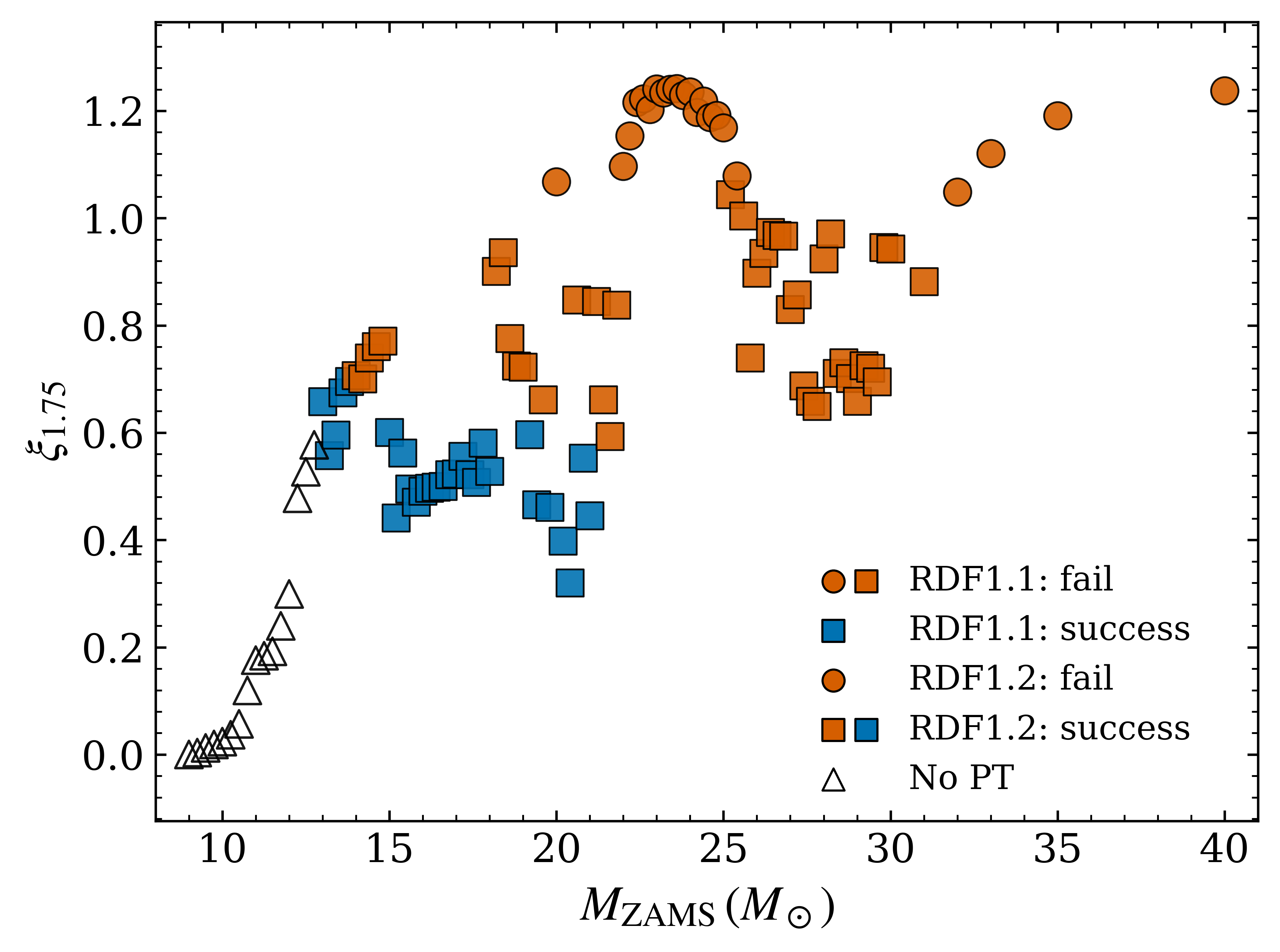}
	\caption{$\xi_{1.75}$ as a function of the initial zero-age main-sequence mass $M_{\rm{ZAMS}}$. Colors distinguish the simulation results obtained with the RDF1.1 EOS (vermilion: fail; blue: success), while marker shapes indicate the results for the RDF1.2 EOS (circle: fail; square: success). Hollow triangles denote models in which the mass is too low to trigger PT.}
	\label{fig:3}
\end{figure}

\begin{figure*}[htbp]
    \includegraphics[width=1.\textwidth]{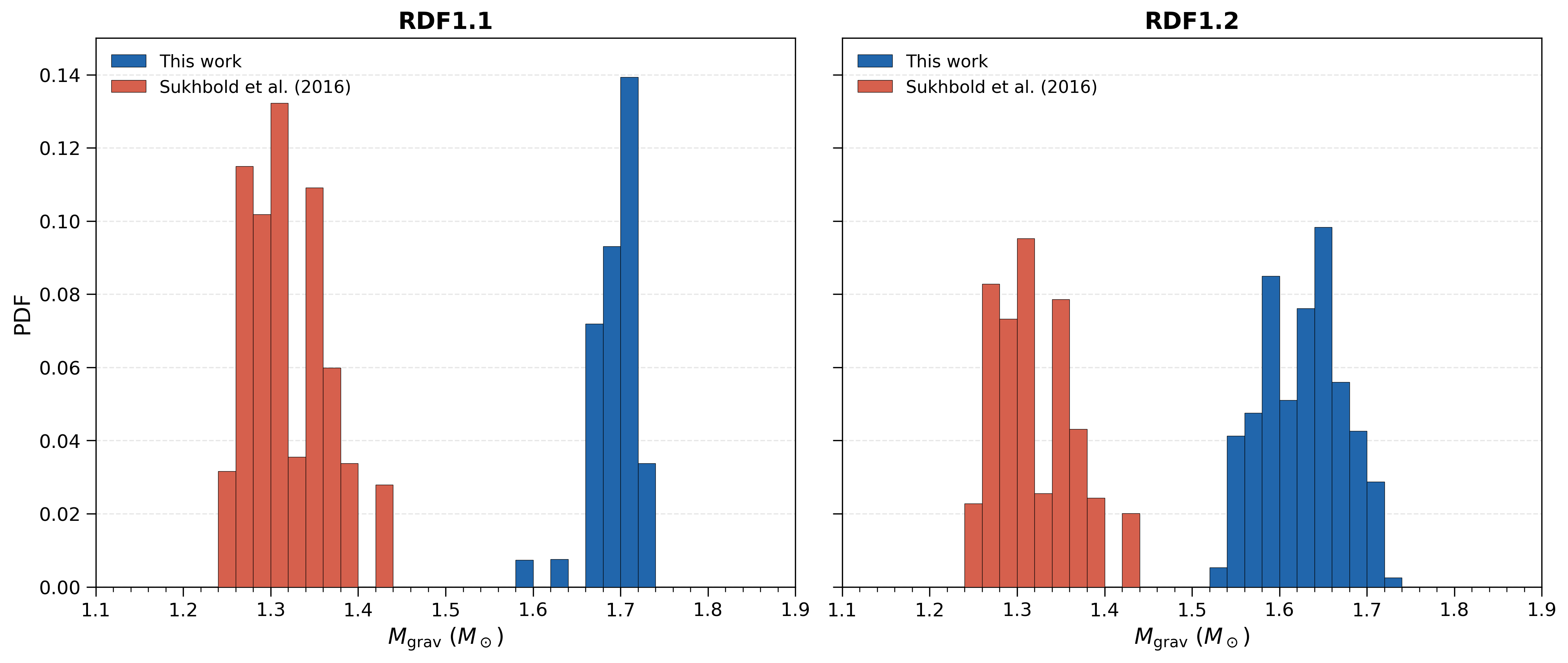}
    \caption{ Probability density function (PDF) of the neutron star gravitational mass distribution for the RDF1.1 (left panel) and RDF1.2 (right panel) EOS. The distribution is derived from stellar population synthesis employing an IMF (initial mass function) and is normalized. The dark blue bars represent the simulation results from this work. For low-compactness progenitors ($M_{\text{ZAMS}}$ $<$ 13 $M_{\odot}$), the mass is too low to trigger PT. We therefore adopt the simulation data of \citet{2016ApJ...821...38S} (red bars) for these mass ranges to reconstruct the complete mass distribution.}
    \label{fig:4}
\end{figure*}

Fig.~\ref{fig:3} shows the explosion simulation results of pre-supernova models with $M_{\rm ZAMS}$ ranging from 13 to 40 $M_{\odot}$ (the complete data for 13--120 $M_\odot$ are listed in Appendix~\ref{app:B}). The horizontal axis represents the initial mass $M_{\rm ZAMS}$, and the vertical axis denotes the compactness parameter $\xi_{1.75}$.

In Fig.~\ref{fig:3}, simulation results using the RDF1.1 EOS are distinguished by color (crimson: failure; blue: success), while those using the RDF1.2 EOS are differentiated by shape (circle: failure; square: success). For pro-supernova models with $M_{\text{ZAMS}} < 13$ $M_\odot$, since the mass within the simulation domain is lower than the phase transition onset mass $M_{\text{onset}}$ of the RDF1.1 and RDF1.2 EOS at $s=3k_{\rm B}$ and $Y_e=0.3$, no PT occurs in the simulations. For the two EOSs, the critical compactness range leading to explosion failure is approximately $\xi_{1.75} \sim 0.59$--$0.69$ for RDF1.1, whereas for RDF1.2, this range shifts upward to $\xi_{1.75} \sim 1.04$--$1.07$. The RDF1.2 enables successful explosions over a broader initial mass range, requiring a greater compactness to result in a failed supernova (black hole formation) \cite{2015MNRAS.450.3289G}. This is consistent with the physical mechanism revealed in Section \ref{sec:3.1}, namely that a lower phase-transition onset density facilitates an earlier triggering of the second shock, thereby providing effective core support.

Fig.~\ref{fig:4} shows the simulated gravitational mass distribution of NSs, with the left and right panels corresponding to the RDF1.1 and RDF1.2 EOSs, respectively. For models in Fig.~\ref{fig:3} where the mass within the simulation domain is insufficient to trigger a PT, we supplement the data using the neutron star masses predicted by \citet{2016ApJ...821...38S}. The distribution is weighted by the stellar population using the initial mass function (IMF, $dN/dM \propto M^{-2.35}$) \cite{1955ApJ...121..161S,1993MNRAS.262..545K} and is normalized. The mass distribution is constructed from two separate
simulation sets. The NS masses for progenitors below the PT threshold cluster around $\sim 1.4\,M_\odot$ (from \citet{2016ApJ...821...38S}). The high mass component, arising directly from the PT simulations in this work, forms a pronounced peak: for RDF1.1, it is centered around $1.70\,M_\odot$, whereas for RDF1.2, it is concentrated at $1.55$--$1.7\,M_\odot$. This difference in the high-mass location between the two EOSs similarly arises from the effect of the phase-transition onset timing on accretion truncation, as discussed in Section \ref{sec:3.1}.

In summary, these results clearly demonstrate that the PT mechanism alone can naturally generate a higher mass peak at about 1.70 $M_\odot$ (RDF1.1) in the birth mass distribution of NSs. If one further accounts for accretion processes in binary systems, where most neutron stars reside, the position of this peak could shift upwards, potentially aligning with the observed high-mass peak at about 1.80 $M_\odot$. Furthermore, physical processes not included in our simulations (such as neutrino FC, neutrino driven turbulence and convection, and fallback accretion) may reshape the explosion boundaries, thereby further refining the quantitative details of the mass distribution.

\subsection{PT induced mass clustering and the higher mass peak}\label{sec:3.3}
\begin{figure*}[htbp]
    \includegraphics[width=1.\textwidth]{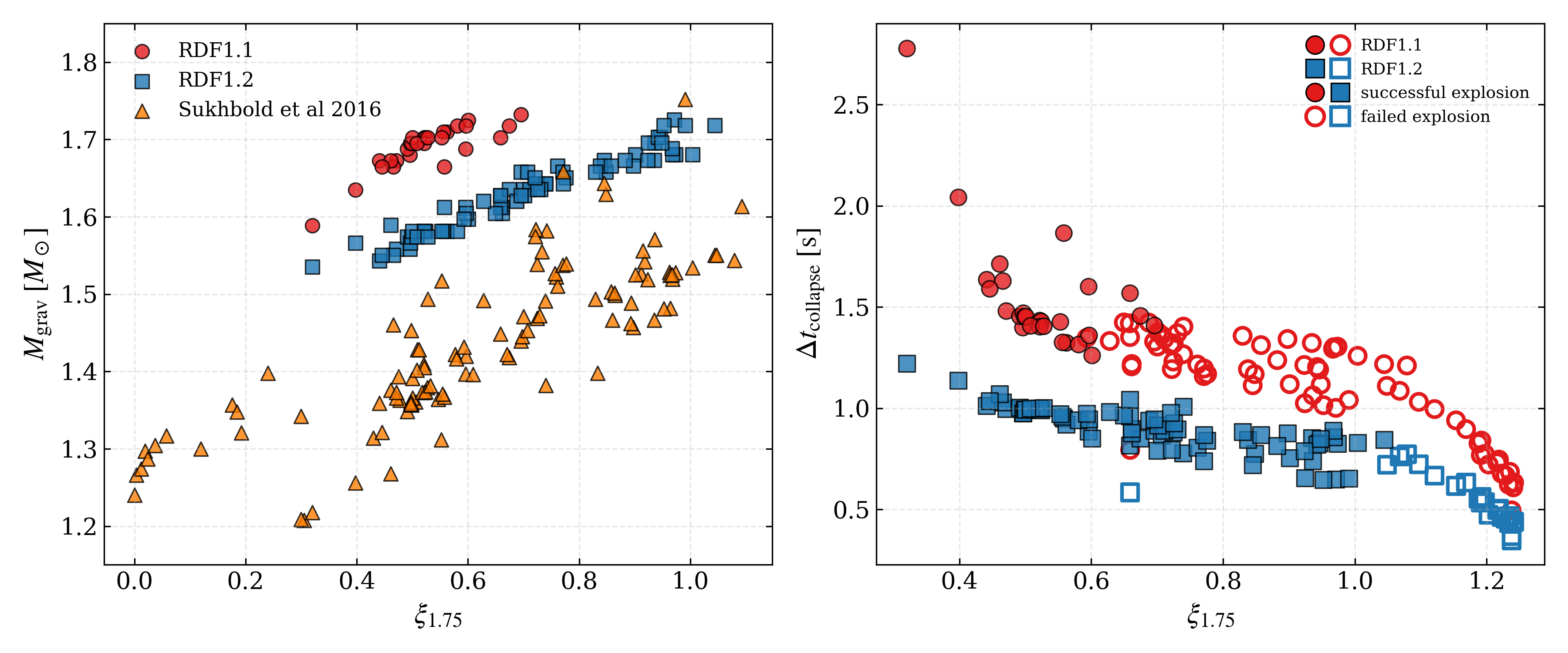}
    \caption{The left panel shows the gravitational mass $M_{grav}$ of the cold NS as a function of the compactness parameter $\xi_{1.75}$. Red circles and blue squares represent the simulation results obtained using the EOS RDF1.1 and RDF1.2, respectively. Orange triangles denote the results from \citet{2016ApJ...821...38S} (calibrated with Z9.6 and N20).  The right panel displays the time interval $\Delta t_{\mathrm{collapse}}$ from shock bounce to second collapse versus the compactness parameter $\xi_{1.75}$. The symbols follow the same convention as in the left panel, where solid symbols indicate successful explosions and hollow symbols indicate collapse to a black hole.}
    \label{fig:5}
\end{figure*}

The left panel of Fig.~\ref{fig:5} displays the final gravitational mass of the NS as a function of the pre-supernova compactness parameter $\xi_{1.75}$. Compared with the simulation results of \citet{2016ApJ...821...38S}, our simulations exhibit two distinct features: First, for models with low to intermediate compactness, the NS masses produced in our simulations are systematically higher; Second, in contrast to the broad mass distribution found in standard models, our results demonstrate a high degree of concentration across different compactness values, constraining the NS masses to a narrow high mass interval (concentrated at $1.66\,M_{\odot}$--$1.72\,M_{\odot}$ for RDF1.1 and $1.55\,M_{\odot}$--$1.70\,M_{\odot}$ for RDF1.2). This systematic mass increase and the clustering effect of the distribution reveal that the PT mechanism plays a critical constraining role in determining the final NS mass.

To further elucidate the physical picture of this mechanism, the right panel of  Fig.~\ref{fig:5} displays the time interval from shock bounce to second collapse, $\Delta t_{\text{collapse}}$, as a function of the compactness parameter $\xi_{1.75}$. The results reveal a distinct anti-correlation between $\Delta t_{\text{collapse}}$ and $\xi_{1.75}$: a lower pre-supernova compactness leads to a later onset of the second collapse. This phenomenon unveils a novel formation channel for relatively more massive NSs driven by the PT. In the standard one-dimensional neutrino-driven explosion paradigm, low-compactness progenitors typically undergo early explosions due to their lower core binding energy, thereby forming less massive NSs ($\sim 1.4\,M_\odot$). However, in our PT incorporated simulations, for low-compactness models where the initial shock fails to trigger a successful explosion, the accretion process persists, leading to a steady growth in the PNS mass. Because of the low compactness, a prolonged accretion phase is required for the central density of the PNS to reach the critical threshold for the PT; during this period, the PNS accumulates substantial material prior to the triggering of the phase transition. Ultimately, the second shock generated by the PT effectively truncates the accretion flow, leaving behind a relatively more massive neutron star remnant.

The physical origin of this mass clustering lies in the fact that, although the precise onset density exhibits sensitivity to the temperature T and the electron fraction Y$_e$, the phase transition boundary itself constitutes a fixed geometric structure within the thermodynamic state space, determined solely by the microscopic strong interaction parameters of the EOS \cite{2021PhRvD.103b3001B,2024ApJ...964..143K}. Consequently, for a specific EOS, the PNS mass at the onset of second collapse is constrained to a relatively narrow range. Therefore, for models resulting in successful explosions, the final neutron star masses inevitably cluster around a characteristic mass corresponding to the second collapse, thereby generating a prominent peak in the overall mass distribution.

Given that the IMF heavily favors low-mass stars, low compactness pre-supernovae dominate numerically on a cosmological scale. Consequently, the PT mechanism provides a novel channel for these abundant low-compactness progenitors to form more massive NSs, offering a natural physical explanation for the high-mass peak in the observed bimodal NS mass distribution. Conversely, precise future observations of the NS mass distribution will serve as a crucial probe for constraining the properties of the hadron-quark PT in dense matter at supranuclear densities.
\begin{figure*}[htbp]
    \includegraphics[width=1.\textwidth]{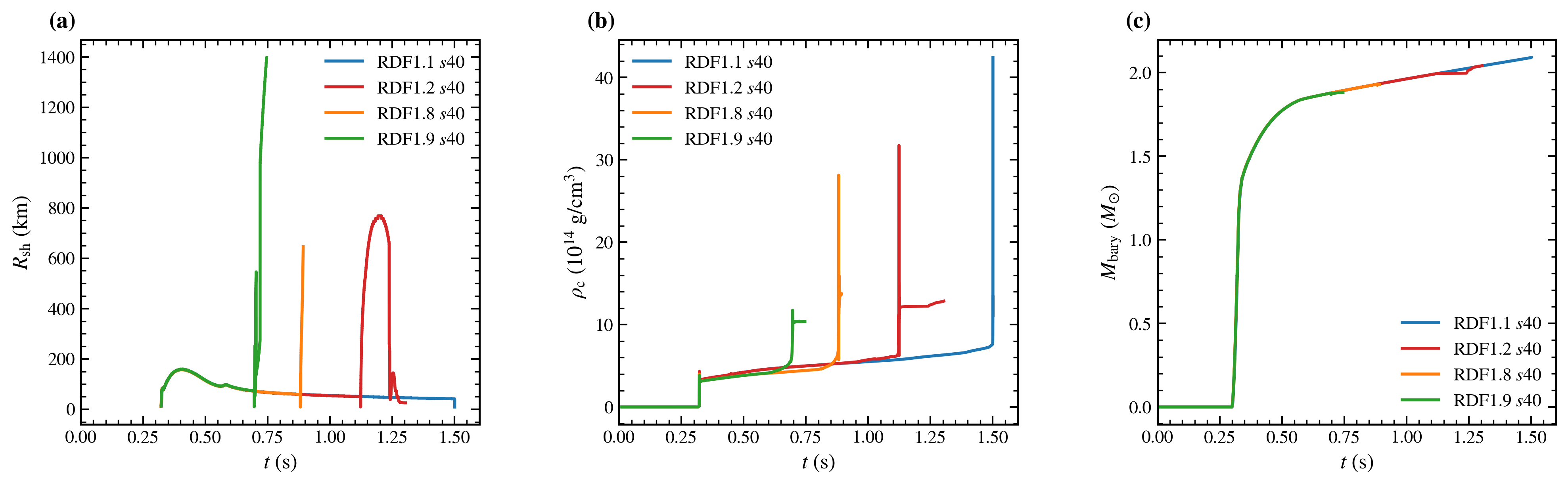}
    \caption{Time evolution of shock radius (panels a), central density (panels b), and PNS mass (defined as the baryonic mass enclosed within the $\rho = 10^{11}\ \mathrm{g\,cm^{-3}}$ contour; panels c) for CCSNe simulations with initial stellar masses of 40 $M_{\odot}$ \citep{2002RvMP...74.1015W}.  Blue lines denote results from simulations using the RDF1.1 equation of state, red lines correspond to RDF1.2, yellow lines represent RDF1.8, and green lines indicate RDF1.9.}
    \label{fig:appendix-1}
\end{figure*}
\section{Conclusion} \label{subsec:Conclusion}
In this study, we systematically investigate the impact of the hadron-quark PT on the explosion dynamics of CCSNe and the NS mass distribution. Utilizing the 1D general relativistic hydrodynamics code GR1D, combined with hybrid EOSs incorporating a first-order phase transition (RDF1.1 and RDF1.2), we perform extensive numerical simulations of massive progenitor models spanning 9 to 120 $M_{\odot}$, and thereby derive a series of important conclusions.

First, by comparing the simulation results obtained with the two hybrid EOSs, RDF1.1 and RDF1.2, we find that the differences between the two EOSs lead to markedly distinct supernova outcomes. These differences are primarily attributed to the different PT onset densities, which exert a threefold impact on the final fate of the supernova. In the low-compactness regime, an earlier PT (RDF1.2) truncates the accretion flow by prematurely triggering the second shock, thereby producing a less massive NS. In the critical-compactness regime, an earlier PT ensures that the PNS explodes successfully before its mass exceeds the maximum support limit of the quark phase, directly dictating the success or failure of the explosion. In the high-compactness regime, however, the extreme mass accretion rate renders the quark phase unable to support the core, meaning a lower onset density conversely accelerates black hole formation.

Second, in the conventional core-collapse scenario, low-compactness progenitors typically yield NSs with masses of $\sim 1.4\,M_\odot$. When the hadron–quark PT is taken into account, the secondary shock triggered by the transition truncates the accretion flow, allowing the proto–neutron star mass to grow to about $1.55\,M_{\odot}$--$1.7\,M_{\odot}$ (specifically, the compact remnant mass peaks near 1.7 $M_{\odot}$ for model RDF1.1, while it spans $1.55\,M_{\odot}$--$1.7\,M_{\odot}$ for model RDF1.2).  If further mass accumulation via binary companion accretion or late-stage fallback is included, this mechanism offers a viable pathway toward explaining the high-mass peak in the observed neutron star mass distribution.

Finally, the physical origin of this higher mass peak is twofold. First, unlike previous studies that primarily investigated whether high compactness progenitors can explode to form massive NSs, we find that low compactness pro-supernova can also yield relatively more massive NSs. Specifically, if the first shock fails to explode, these systems continue to accrete until the formation of a second shock. This second shock subsequently triggers the supernova explosion, leading to the formation of relatively more massive NSs. Second, for a given hybrid EOS, as the PT is triggered when the PNS core evolves to a relatively fixed critical phase boundary, the remnant masses resulting from PT driven explosions inevitably cluster around a characteristic mass associated with the second collapse.

Although our 1D simulation framework omits multidimensional physical effects such as neutrino FC, neutrino-driven turbulence and convection, post-explosion fallback accretion, and binary interactions, the conclusion regarding the new formation channel for the more massive neutron stars and the clustering effect induced by the phase transition remains robust. This robustness stems from the fact that the underlying mechanism is intrinsically determined by the properties of the equation of state incorporating the phase transition, rather than relying on fluid dynamical instabilities. In the future, precise observational constraints on the NS mass distribution will, in turn, serve as an important astrophysical probe for exploring the properties of the PT in high-density nuclear matter.

\section*{Acknowledgements}
This work was supported by the National Natural Science Foundation of China (Grants Nos. 12373038, 12541303 and 12563007), the Natural Science Foundation of Xinjiang (Grant No. 2022D01D85), the China Manned Space Program (Grant No.CMS-CSST-2025-A15) and the Graduate Student Research (Practice) Innovation Project of Xinjiang Uygur Autonomous Region (Grant No. XJ2026G015).

\appendix

\section{Simulation comparison} \label{app:A}
Fig.~\ref{fig:appendix-1} displays the simulation results for the s40 pre-supernova progenitor model utilizing various equations of state. The simulation employing the RDF1.1 EOS collapses promptly into a black hole, with a PNS baryonic mass of 2.09 $M_{\odot}$ at the time of termination. The RDF1.2 EOS model undergoes a delayed collapse into a black hole, yielding a remnant mass of 2.04 $M_{\odot}$. In contrast, the RDF1.8 and RDF1.9 models successfully trigger explosions, resulting in NS baryonic masses of 1.93 $M_{\odot}$ and 1.88 $M_{\odot}$, respectively. For comparison, \citet{2024ApJ...964..143K} reported masses of 2.10 $M_{\odot}$, 2.02 $M_{\odot}$, 1.95 $M_{\odot}$, and 1.88 $M_{\odot}$ for the corresponding EOSs in their Table 4. The discrepancies between our results and those of \citet{2024ApJ...964..143K} are consistently within 0.02 $M_{\odot}$. This deviation lies well within the systematic uncertainties expected between different supernova simulation codes (see the benchmark comparison in \citet{2018JPhG...45j4001O}), thereby robustly validating the reliability of our simulation results.
\section{Table} \label{app:B}
\begin{table*}[htbp]
\centering
\caption{Summary of CCSNe simulation results for two EOSs: RDF1.1 (left) and RDF1.2 (right). $M_{\rm ZAMS}$ is the zero-age main-sequence mass, $t_{\rm bounce}$ and $t_{\rm collapse}$ are the shock bounce and second collapse times, $M_{\rm bary}$ is the enclosed PNS remnant baryon mass at a density of $10^{11}~{\rm g~cm^{-3}}$ at the end of the SN simulations, and $M_{\rm grav}$ is the cold gravitational mass of the NS, "---" indicate explosion failure.}
\label{tab:results}
\scriptsize
\renewcommand{\arraystretch}{0.55}
\begin{ruledtabular}
\begin{tabular}{ccccc|ccccc}
\multicolumn{5}{c|}{\textbf{EOS: RDF1.1}} & \multicolumn{5}{c}{\textbf{EOS: RDF1.2}} \\
\hline
$M_{\rm ZAMS}$ & $t_{\rm bounce}$ & $t_{\rm collapse}$ & $M_{\rm bary}$ & $M_{\rm grav}$ & $M_{\rm ZAMS}$ & $t_{\rm bounce}$ & $t_{\rm collapse}$ & $M_{\rm bary}$ & $M_{\rm grav}$ \\
($M_\odot$) & (s) & (s) & ($M_\odot$) & ($M_\odot$) & ($M_\odot$) & (s) & (s) & ($M_\odot$) & ($M_\odot$) \\
\hline
13.0 & 0.265 & 1.835 & 1.95 & 1.70 & 13.0 & 0.264 & 1.080 & 1.85 & 1.63 \\
13.2 & 0.269 & 2.135 & 1.90 & 1.67 & 13.2 & 0.270 & 1.220 & 1.83 & 1.61 \\
13.4 & 0.264 & 1.866 & 1.93 & 1.69 & 13.4 & 0.264 & 1.150 & 1.83 & 1.61 \\
13.6 & 0.266 & 1.723 & 1.97 & 1.72 & 13.6 & 0.265 & 1.115 & 1.86 & 1.64 \\
13.8 & 0.295 & 1.705 & 1.99 & 1.73 & 13.8 & 0.296 & 1.185 & 1.89 & 1.66 \\
14.0 & 0.290 & 1.650 & 1.99 & --- & 14.0 & 0.290 & 1.155 & 1.89 & 1.66 \\
14.2 & 0.285 & 1.590 & 1.99 & --- & 14.2 & 0.285 & 1.075 & 1.86 & 1.64 \\
14.4 & 0.273 & 1.540 & 2.01 & --- & 14.4 & 0.273 & 1.050 & 1.87 & 1.64 \\
14.6 & 0.284 & 1.500 & 2.02 & --- & 14.6 & 0.284 & 1.090 & 1.90 & 1.67 \\
14.8 & 0.287 & 1.445 & 2.04 & --- & 14.8 & 0.287 & 1.025 & 1.89 & 1.66 \\
15.0 & 0.224 & 1.485 & 1.98 & 1.72 & 15.0 & 0.224 & 1.075 & 1.81 & 1.60 \\
15.2 & 0.223 & 1.860 & 1.91 & 1.67 & 15.2 & 0.223 & 1.235 & 1.74 & 1.54 \\
15.4 & 0.226 & 1.550 & 1.96 & 1.71 & 15.4 & 0.226 & 1.145 & 1.79 & 1.58 \\
15.6 & 0.230 & 1.630 & 1.92 & 1.68 & 15.6 & 0.229 & 1.205 & 1.76 & 1.56 \\
15.8 & 0.238 & 1.720 & 1.91 & 1.67 & 15.8 & 0.238 & 1.235 & 1.76 & 1.56 \\
16.0 & 0.246 & 1.700 & 1.93 & 1.69 & 16.0 & 0.245 & 1.250 & 1.78 & 1.57 \\
16.2 & 0.244 & 1.715 & 1.94 & 1.70 & 16.2 & 0.244 & 1.220 & 1.77 & 1.57 \\
16.4 & 0.246 & 1.700 & 1.94 & 1.70 & 16.4 & 0.246 & 1.240 & 1.78 & 1.57 \\
16.6 & 0.252 & 1.705 & 1.95 & 1.70 & 16.6 & 0.251 & 1.245 & 1.79 & 1.58 \\
16.8 & 0.256 & 1.690 & 1.95 & 1.70 & 16.8 & 0.255 & 1.250 & 1.79 & 1.58 \\
17.0 & 0.256 & 1.685 & 1.95 & 1.70 & 17.0 & 0.255 & 1.245 & 1.79 & 1.58 \\
17.2 & 0.238 & 1.565 & 1.96 & 1.71 & 17.2 & 0.237 & 1.195 & 1.79 & 1.58 \\
17.4 & 0.246 & 1.650 & 1.94 & 1.70 & 17.4 & 0.245 & 1.245 & 1.79 & 1.58 \\
17.6 & 0.251 & 1.660 & 1.94 & 1.70 & 17.6 & 0.251 & 1.250 & 1.78 & 1.57 \\
17.8 & 0.235 & 1.550 & 1.97 & 1.72 & 17.8 & 0.234 & 1.175 & 1.79 & 1.58 \\
18.0 & 0.235 & 1.640 & 1.95 & 1.70 & 18.0 & 0.234 & 1.235 & 1.78 & 1.57 \\
18.2 & 0.301 & 1.420 & 2.07 & --- & 18.2 & 0.301 & 1.055 & 1.92 & 1.68 \\
18.4 & 0.297 & 1.360 & 2.07 & --- & 18.4 & 0.296 & 1.035 & 1.94 & 1.70 \\
18.6 & 0.291 & 1.460 & 2.05 & --- & 18.6 & 0.291 & 1.130 & 1.88 & 1.65 \\
18.8 & 0.280 & 1.510 & 2.03 & --- & 18.8 & 0.279 & 1.160 & 1.87 & 1.64 \\
19.0 & 0.277 & 1.470 & 2.02 & --- & 19.0 & 0.276 & 1.070 & 1.87 & 1.64 \\
19.2 & 0.260 & 1.620 & 1.97 & 1.72 & 19.2 & 0.259 & 1.205 & 1.82 & 1.60 \\
19.4 & 0.215 & 1.845 & 1.90 & 1.67 & 19.4 & 0.215 & 1.245 & 1.75 & 1.55 \\
19.6 & 0.263 & 1.470 & 2.00 & --- & 19.6 & 0.263 & 1.140 & 1.83 & 1.61 \\
19.8 & 0.291 & 2.005 & 1.91 & 1.67 & 19.8 & 0.291 & 1.360 & 1.80 & 1.59 \\
20.0 & 0.319 & 1.405 & 2.10 & --- & 20.0 & 0.318 & 1.080 & 1.97 & --- \\
20.2 & 0.293 & 2.335 & 1.86 & 1.63 & 20.2 & 0.293 & 1.430 & 1.77 & 1.57 \\
20.4 & 0.286 & 3.065 & 1.80 & 1.59 & 20.4 & 0.285 & 1.505 & 1.73 & 1.54 \\
20.6 & 0.267 & 1.435 & 2.04 & --- & 20.6 & 0.266 & 1.040 & 1.89 & 1.66 \\
20.8 & 0.249 & 1.675 & 1.95 & 1.70 & 20.8 & 0.248 & 1.220 & 1.79 & 1.58 \\
21.0 & 0.254 & 1.845 & 1.90 & 1.67 & 21.0 & 0.254 & 1.290 & 1.75 & 1.55 \\
21.2 & 0.272 & 1.385 & 2.05 & --- & 21.2 & 0.271 & 0.990 & 1.91 & 1.67 \\
21.4 & 0.237 & 1.455 & 1.99 & --- & 21.4 & 0.236 & 1.135 & 1.82 & 1.60 \\
21.6 & 0.273 & 1.620 & 1.98 & --- & 21.6 & 0.273 & 1.245 & 1.81 & 1.60 \\
21.8 & 0.302 & 1.495 & 2.06 & --- & 21.8 & 0.302 & 1.145 & 1.90 & 1.67 \\
22.0 & 0.324 & 1.355 & 2.12 & --- & 22.0 & 0.323 & 1.045 & 2.00 & --- \\
22.2 & 0.335 & 1.275 & 2.16 & --- & 22.2 & 0.334 & 0.950 & 2.05 & --- \\
22.4 & 0.363 & 1.095 & 2.24 & --- & 22.4 & 0.362 & 0.860 & 2.14 & --- \\
22.6 & 0.368 & 1.045 & 2.27 & --- & 22.6 & 0.367 & 0.830 & 2.17 & --- \\
22.8 & 0.373 & 1.095 & 2.24 & --- & 22.8 & 0.372 & 0.845 & 2.16 & --- \\
23.0 & 0.369 & 1.000 & 2.28 & --- & 23.0 & 0.367 & 0.805 & 2.18 & --- \\
23.2 & 0.368 & 0.990 & 2.28 & --- & 23.2 & 0.366 & 0.800 & 2.19 & --- \\
23.4 & 0.366 & 0.975 & 2.27 & --- & 23.4 & 0.365 & 0.805 & 2.18 & --- \\
23.6 & 0.368 & 1.000 & 2.28 & --- & 23.6 & 0.366 & 0.805 & 2.18 & --- \\
23.8 & 0.352 & 1.020 & 2.27 & --- & 23.8 & 0.351 & 0.805 & 2.17 & --- \\
24.0 & 0.349 & 1.035 & 2.26 & --- & 24.0 & 0.348 & 0.815 & 2.16 & --- \\
24.2 & 0.346 & 1.120 & 2.21 & --- & 24.2 & 0.344 & 0.880 & 2.11 & --- \\
24.4 & 0.349 & 1.095 & 2.24 & --- & 24.4 & 0.348 & 0.850 & 2.13 & --- \\
24.6 & 0.343 & 1.170 & 2.20 & --- & 24.6 & 0.342 & 0.895 & 2.09 & --- \\
24.8 & 0.350 & 1.190 & 2.20 & --- & 24.8 & 0.349 & 0.910 & 2.09 & --- \\
25.0 & 0.339 & 1.235 & 2.17 & --- & 25.0 & 0.337 & 0.970 & 2.07 & --- \\
25.2 & 0.327 & 1.545 & 2.07 & --- & 25.2 & 0.327 & 1.170 & 1.97 & 1.72 \\
25.4 & 0.324 & 1.535 & 2.09 & --- & 25.4 & 0.323 & 1.095 & 1.98 & --- \\
25.6 & 0.306 & 1.565 & 2.05 & --- & 25.6 & 0.305 & 1.135 & 1.92 & 1.68 \\
25.8 & 0.317 & 1.720 & 1.99 & --- & 25.8 & 0.317 & 1.325 & 1.87 & 1.64 \\
26.0 & 0.308 & 1.650 & 2.02 & --- & 26.0 & 0.307 & 1.185 & 1.90 & 1.67 \\
26.2 & 0.313 & 1.635 & 2.03 & --- & 26.2 & 0.312 & 1.165 & 1.91 & 1.67 \\
26.4 & 0.311 & 1.615 & 2.04 & --- & 26.4 & 0.310 & 1.135 & 1.92 & 1.68 \\
26.6 & 0.308 & 1.610 & 2.04 & --- & 26.6 & 0.308 & 1.165 & 1.92 & 1.68 \\
26.8 & 0.311 & 1.605 & 2.04 & --- & 26.8 & 0.310 & 1.200 & 1.93 & 1.69 \\
27.0 & 0.297 & 1.655 & 2.01 & --- & 27.0 & 0.297 & 1.180 & 1.89 & 1.66 \\
27.2 & 0.303 & 1.615 & 2.01 & --- & 27.2 & 0.303 & 1.170 & 1.90 & 1.67 \\
27.4 & 0.277 & 1.700 & 1.97 & --- & 27.4 & 0.276 & 1.215 & 1.84 & 1.62 \\
27.6 & 0.276 & 1.695 & 1.96 & --- & 27.6 & 0.275 & 1.235 & 1.83 & 1.61 \\
27.8 & 0.276 & 1.700 & 1.96 & --- & 27.8 & 0.276 & 1.240 & 1.82 & 1.60 \\
28.0 & 0.266 & 1.290 & 2.08 & --- & 28.0 & 0.265 & 0.920 & 1.94 & 1.70 \\
28.2 & 0.278 & 1.280 & 2.10 & --- & 28.2 & 0.277 & 0.925 & 1.98 & 1.73 \\
28.4 & 0.262 & 1.605 & 1.99 & --- & 28.4 & 0.261 & 1.150 & 1.86 & 1.64 \\
28.6 & 0.274 & 1.645 & 2.00 & --- & 28.6 & 0.273 & 1.170 & 1.86 & 1.64 \\
28.8 & 0.270 & 1.645 & 1.98 & --- & 28.8 & 0.269 & 1.185 & 1.85 & 1.63 \\
29.0 & 0.274 & 1.625 & 1.99 & --- & 29.0 & 0.273 & 1.315 & 1.85 & 1.63 \\
29.2 & 0.275 & 1.595 & 2.01 & --- & 29.2 & 0.275 & 1.200 & 1.86 & 1.64 \\
29.4 & 0.284 & 1.595 & 2.01 & --- & 29.4 & 0.283 & 1.260 & 1.88 & 1.65 \\
29.6 & 0.281 & 1.610 & 2.00 & --- & 29.6 & 0.280 & 1.225 & 1.85 & 1.63 \\
29.8 & 0.299 & 1.490 & 2.07 & --- & 29.8 & 0.298 & 1.125 & 1.95 & 1.70 \\
30.0 & 0.302 & 1.505 & 2.07 & --- & 30.0 & 0.300 & 1.120 & 1.95 & 1.70 \\
31.0 & 0.298 & 1.535 & 2.04 & --- & 31.0 & 0.297 & 1.110 & 1.91 & 1.67 \\
32.0 & 0.315 & 1.425 & 2.09 & --- & 32.0 & 0.314 & 1.035 & 1.98 & --- \\
33.0 & 0.329 & 1.325 & 2.13 & --- & 33.0 & 0.328 & 0.995 & 2.02 & --- \\
35.0 & 0.347 & 1.115 & 2.21 & --- & 35.0 & 0.345 & 0.880 & 2.11 & --- \\
40.0 & 0.394 & 0.850 & 2.37 & --- & 40.0 & 0.392 & 0.740 & 2.27 & --- \\
45.0 & 0.385 & 0.880 & 2.35 & --- & 45.0 & 0.383 & 0.750 & 2.24 & --- \\
50.0 & 0.297 & 1.510 & 2.04 & --- & 50.0 & 0.296 & 1.085 & 1.91 & 1.67 \\
55.0 & 2.096 & 2.890 & 2.17 & --- & 55.0 & 2.096 & 2.680 & 2.06 & --- \\
60.0 & 0.278 & 1.610 & 1.99 & --- & 60.0 & 0.277 & 1.260 & 1.84 & 1.62 \\
70.0 & 0.308 & 1.425 & 2.07 & --- & 70.0 & 0.307 & 1.155 & 1.94 & 1.70 \\
80.0 & 0.299 & 1.495 & 2.04 & --- & 80.0 & 0.298 & 1.165 & 1.87 & 1.64 \\
100.0 & 0.294 & 1.335 & 2.11 & --- & 100.0 & 0.293 & 0.945 & 1.97 & 1.72 \\
120.0 & 0.280 & 1.295 & 2.10 & --- & 120.0 & 0.279 & 0.925 & 1.97 & 1.72 \\
\hline
\end{tabular}
\end{ruledtabular}
\end{table*}
\bibliography{ms}
\end{document}